\RequirePackage{lineno}
\documentclass[10pt,aps,prb,final,floatfix,notitlepage,twocolumn,superscriptaddress]{iopart}
\usepackage{graphicx}
\usepackage{epsfig}
\usepackage[scriptsize,bf]{caption}
\usepackage{subfig}
\begin{document}               % plus the \end{document} command at the end.
%\modulolinenumbers[5]
%\linenumbers
\title[Nano--particle enhanced radiation treatment]{Beam Energy Considerations for Gold Nano--Particle Enhanced Radiation Treatment}
\author{F. Van den Heuvel Ph.D.$^*$$^1$, Jean--Pierre Locquet Ph.D.$^2$ and  S. Nuyts M.D. Ph.D.$^1$}
\address{Dept of Experimental Radiotherapy University of Leuven, Leuven, Belgium}
\ead{($^*$)frank.vandenheuvel@med.kuleuven.be}
\address{Solid State Physics and Magnetism Section University of Leuven, Leuven, Belgium}
\date{\today}
\begin{abstract}
{\bf Purpose:}
A novel approach using nano technology enhanced
radiation modalities is investigated. The proposed methodology uses
antibodies labeled with organically inert metals with a high atomic
number. Irradiation using photons with energies in the kilo--electron volt
(keV) range show an increase in dose due to a combination of an increase in
photo--electric interactions and a pronounced generation of Auger and/or
Coster--Kr\"onig (A--CK) electrons.
\par
{\bf Methods:}
The dependency of the dose deposition on various factors is investigated
using Monte Carlo simulation models. The factors investigated include:
agent concentration, spectral dependency looking at mono--energetic
sources as well as classical bremsstrahlung sources. The optimization of the energy spectrum 
is performed in terms of physical dose enhancement as well as the dose deposited by Auger and/or 
Coster--Kr\"onig electrons and their biological effectiveness.
\par
{\bf Results:}
A quasi--linear dependency on concentration and an exponential decrease
within the target medium is observed. The maximal dose enhancement is dependent 
on the position of the target in the beam. Apart from irradiation with low photon energies (10 -- 20 keV)
there is no added benefit from the increase in generation of Auger electrons.
Interestingly, a regular 110kVp bremsstrahlung spectrum shows a comparable
enhancement in comparison with the optimized mono--energetic sources. 
\par {\bf Conclusions:} In conclusion we find that the use of nano--particle enhanced shows
promise to be implemented quite easily in regular clinic on a physical
level due to the advantageous properties in classical beams. 
\end{abstract}
\maketitle
\section{Introduction}
Recently, methodologies using monoclonal antibodies that target specific
tumor cells have been used to bring active compounds in the vicinity of
these cells. One  approach uses radioactive compounds of
$\alpha$-- or
$\beta$--emitters\cite{Nayak2007185,Pless2004365,Miederer20081371}.
Alternatively, chemotherapeutic compounds
have been attached to this delivery mechanism.
The use of such approaches is interesting but limited due to the
fact that the therapeutic compound is already active at time of
delivery and during secretion by the body. More in particular with
radioactive compounds an important whole body dose (red marrow dose)
as well as renal toxicity are limiting factors for the efficacy of the
treatment\cite{Otte2002,Schumacher2002,Behr2002,Cybulla2001}.
\par 
It is the goal of this paper to investigate a delivery method that could
potentially have most of the benefits associated with the previously listed
therapeutic modalities \'and has almost none of the disadvantages. Which
means:
\begin{enumerate}
\item Differentiation between malignant and healthy cells.
\item Enhanced effectiveness.
\item Image guidance possibilities.
\item Activation methodology (i.e. Only active were it needs to be active).
\item Large therapeutic window.
\end{enumerate}
Dose enhancement due to the presence of gold nano--particles has been
proposed already both by means of an injectable contrast agent as by the use
of mono--clonal antibodies or other targetted delivery methods.. However,
the enhancement relied on an increased interaction due to the
increased probability of the photo--electric interaction being dependent
on the atomic number at the proposed energies and the specific contribution of
Auger electrons was not investigated\cite{Hainfeld2004,Verhaegen2006,Cho2005,Cho2009}.
All sources used in these studies were spectral sources and/or
brachytherapy sources.  Moreover, proposals to use more sophisticated
photon sources have been put forward, in the hope to maximize the
efficiency of the conversion of the beam energy to deposited energy as
well as generate a high amount of Auger electrons\cite{silver:2899}.  
\par 
In the dose deposition model proposed here a significant part of the
energy is deposited by Auger electrons. There is reason to believe
that Auger electrons deposit their energy more efficiently than those
emanating from Compton or photo--electric effect processes.
The exact mechanism behind this apparant dose enhancement effect is still unclear. A possible cause is
the fact that Auger have a very low energy and
deposit all of the energy within a range comparable to a few cell
diameters. Furthermore, there is a possible change in the stopping power energy dependency at very low energies ($<$10keV), where the Bethe formalism breaks down. 
Alternatively, it could be that on average more than a single Auger electron is being produced, increasing the probability of clustered double strand breaks.
\par
 A number of authors have investigated the biological effects
indirectly and support the notion that Auger electrons indeed have high
LET characteristics\cite{Chen2008,Urashima2006,Balagurumoorthy2008}.
\par
To our knowledge. a systematic study of the impact of different spectral sources on the enhancement and the possible biological enhancement has not been published. 
\section{Methods and Materials}
To perform the planning simulation we used MCNPX (Monte Carlo N-Particle eXtended)\cite{MCNPXref} version 2.7a running on a 
738 node cluster at the University of Leuven. The department of experimental radiotherapy at Leuven is part of the beta--test group for MCNPX. 
The following physical parameters were used during these simulations. 
\begin{description}
\item [Photon energy cutoff:] 1keV.
\item [electron energy cutoff:] 1keV.
\item [EM interaction library:] ENDF/B-VI Release 8 Photoatomic Data 02/07/03
\end{description}
\par For the spectral dose deposition from Auger electrons version 2.7b was used. As the version is only available to use as a single processor binary
it was not possible to run this on the cluster. Therefore if we did not need information on the Auger electrons seperately we chose to use the earlier version.
\subsection{Geometry}
\par
The simulation geometry consisted of a tank filled with water containing three 1mm thick 
cylindrical slabs holding tissue as defined by ORNL report TM8381. One slab was positioned at the surface of the 
tank representing a skin surface. A second slab was
positioned at 5cm distance from the tank surface down stream from the
source. In this slab varying concentrations of gold were added in a homogeneous distribution. Additionally, geometrically identical slabs containing 
tissue, were positioned downstream adjacent to
the structure containing the gold particles. 
The source in this geometry was a plane source (not divergent). The divergence can be introduced depending of the position of
the source in a clinical situation by applying a inverse square rule. 
\begin{figure}[h]
\begin{center}
\includegraphics[width=0.7\columnwidth]{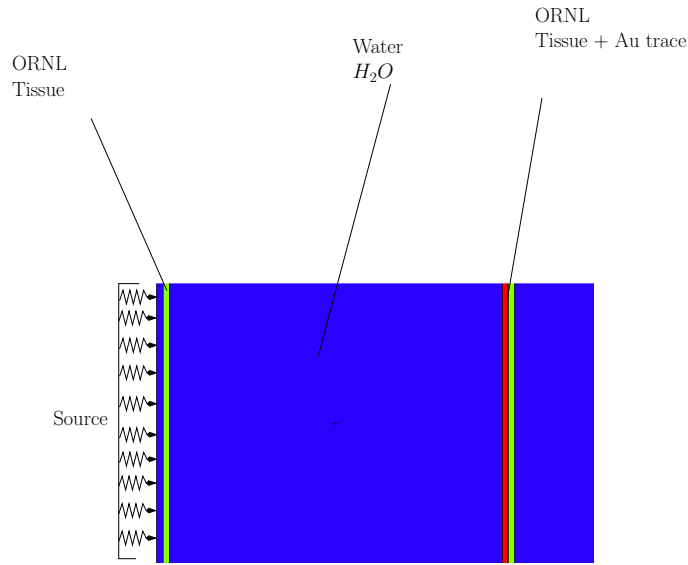}
\end{center}
\caption{\label{geom} The geometry used in this simulation. The geometry
has three volumes where dose can be measured:  A skin section
at the edge of the area, a target volume, and an organ at risk volume
adjacent to the target volume. Alternatively the third volume can be used as a layer of target volume
to study the effect of upstream nano--particles.}
\end{figure}
\subsection{Source}
\par The radiation source is modeled after mono--energetic
radiation sources obtained by Bragg--Gray diffraction of regular X--ray
sources. Although these sources are called mono--energetic, they do
exhibit some spectral spread which was modeled as a normal distribution
with a $\sigma$= 1.5keV. 
\subsection{Simulations}
\subsubsection{Energies}
To investigate the energy dependence of the therapeutic window, we performed the
same simulation with quasi--mono--energetic beams of energies ranging from
10 to  200 keV, with special consideration
to the K$_\alpha$--energy of gold (i.e. 80.67keV). 
Additionally, a broad spectrum beam was investigated. The spectrum was
taken to be identical to that coming from
an Acuity simulator's X--ray tube (Varian Inc.) running at 110kVp. 
\subsubsection{Concentrations}
The medium used in the activated environment was considered to be tissue
as defined in the (Oak Ridge National Labs (ORNL) Report TM-8381). Gold
(Z=79) in natural isotope abundance was added with all other components
diminished to yield a normalized weight. The structures were embedded
in water. The concentrations of gold in one of the structures varied
between 0 and 10\% in steps of 1\%. A concentration of 10\% is highly unlikely, however as reported by 
Verhaegen et al. it is the concentration  of off--the--shelf contrast material and should serve as an upper limit of 
the enhancements achievable with this technique. Furthermore, it is to be expected that once a distribution methodology 
for the gold particles is implemented we are bound to see very heterogeneous concentrations of gold in the irradiated medium,
reaching high concentrations locally.
\par
All concentration related simulations
were performed with the broad spectrum, for reasons made clear
in the results section.
\subsubsection{Auger electrons}
As mentioned above, 
the contribution of the Auger electrons could be estimated by 
tagging the electrons released due to Auger cascades in the cell of interest.
The number of generated Auger electrons could then be linked to the energy of the source.
\par
This to find the optimal energy for Auger electron generation. Below
energies of 1keV MCNPX does not track the electrons and the energy is
considered to be deposited locally. The simulations were performed for all energies as listed above. 
\subsection{Analysis}
\subsubsection{Energy Dependence}
The energy dependence is reviewed for the maximal concentration of
10\%, this because any differences between the energies would be magnified, as well as reduce the 
statistical errors in our monte carlo calculations.  We define three types of dose enhancement:
\begin{description}
\item[Absolute Enhancement ($E_a$):] The ratio of the dose deposited
in the gold--containing structure to the dose deposition in a run with
exact same geometry without gold present.
\item[OAR Enhancement ($E_{OAR}$)]: The ratio of the dose deposited in
the gold--containing structure to the dose deposited in another structure
in the geometry not containing gold, representing an Organ at Risk (OAR).
\item[Skin ratio ($E_{S}$)]: The ratio of the dose deposited in the gold--containing structure to a layer 1mm under the skin. Which is a special case of $E_{OAR}$.
\end{description}
We determined the energy deposited per unit mass by counting the energy deposited by electrons using the MCNPX tally F6:E.
This underestimates the effectively deposited dose slightly as it does not take into account the energy expended to 
generate the photon initiated ionization. However, this ionization is likely not contributing to a biological effect as it is mainly through interaction with the gold atom which does not form part of the cell structure. 
The difference between F6:E and F6:P (Energy transferred by photons, which includes electrons) is about 1\%. However the electron energy deposition is better defined spatially. 
\subsubsection{Concentration}
The results from the concentration study were analyzed in function of absolute enhancement. 
The variation $E_a$ was fit using a linear relationship and a second order polynomial. 
The fit was performed using a minimization of a $\chi^2$--function taking into account the 
simulation errors. 
\subsection{Radiobiological effect}
\begin{figure}[h]
\begin{center}
\includegraphics[width=0.45\columnwidth]{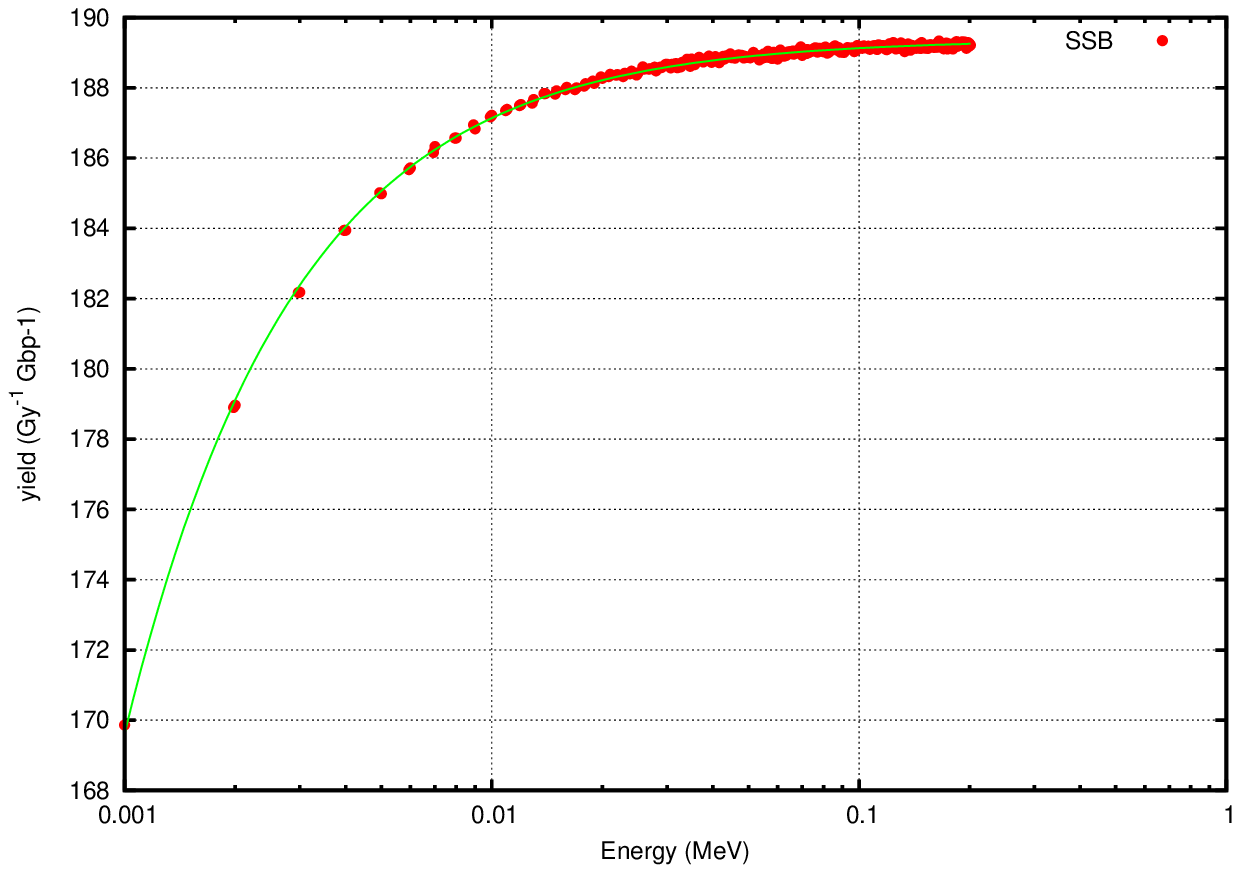}
\includegraphics[width=0.45\columnwidth]{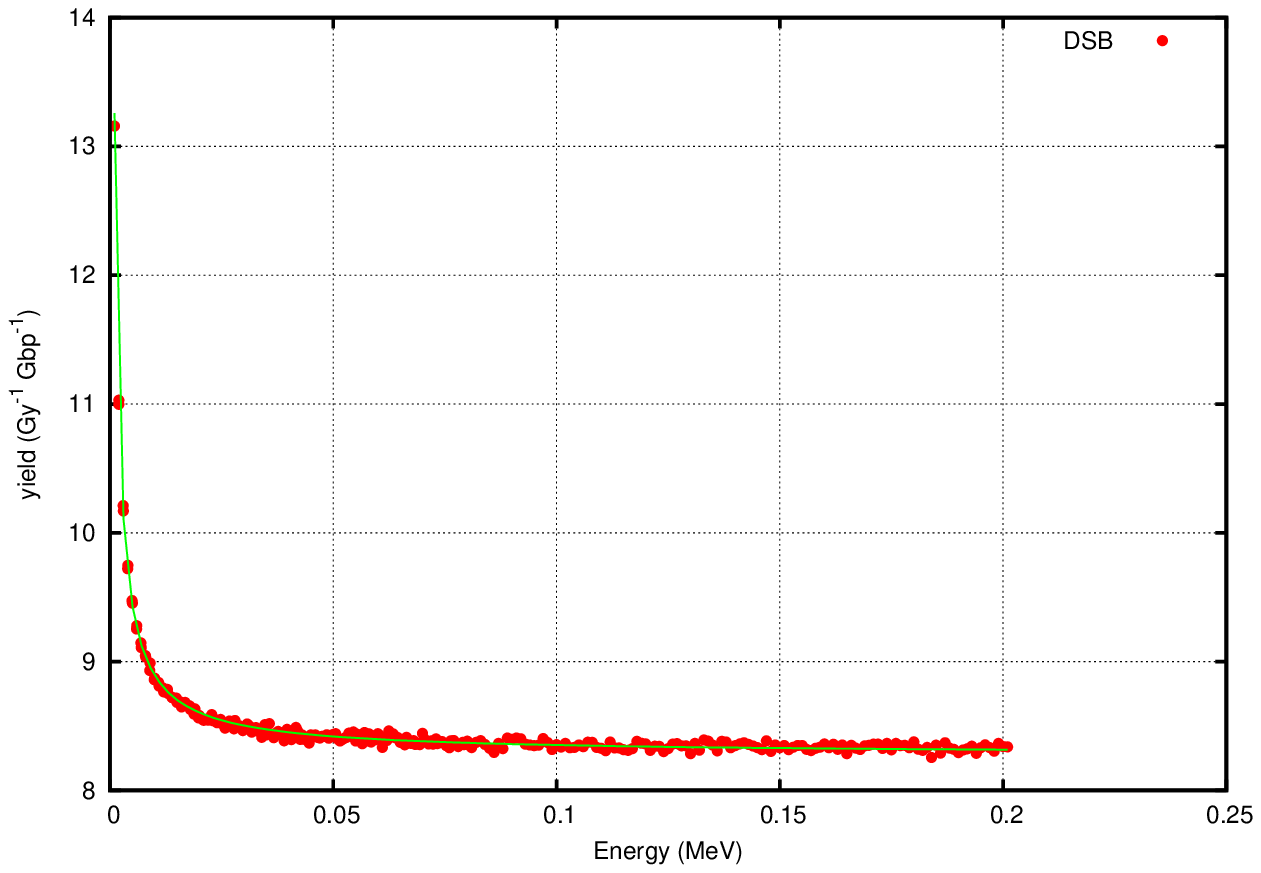}
\end{center}
\caption{\label{DSBfit} The yield of double strand breaks as defined by Nikjoo et al. fit with a function having the general form:
$a+bx^c$, with a, b, and c being parameters. }
\end{figure}

To estimate the relative effect of the change in spectrum and the increase in
the contribution by Auger--electrons we used a fast monte carlo model of
biological damage as proposed by Semenenko and Stewart\cite{ISI:000237044600004}
which has been shown to obtain the same results as track code monte
carlo codes as proposed by Nikjoo et al. \cite{Nikjoo1997}. The approach used
here determines the amount of different damage to a DNA--molecule from
direct ionization as well as through the generation of radicals. It then
produces a yield ($y$) in percentage of the different types of damage
ranging from single strand breaks to complex clustered double strand
breaks for interacting electrons of a given energy ($E$). The methodology used to incorporate this information in our calculations is as follows.
\begin{enumerate}
\item For every energy available in the energy deposition histogram we used the code provided by Semenenko and Stewart to generate the yields of the different 
types of DNA--damage making sure that the lowest energy is included. The yield 
is given as percentage per cell and per Gy ($\%\textrm{cell}^{-1}\textrm{Gy}^{-1}$). 
\item The data was fit as a function of energy of the electron using an equation of the form:
\begin{equation}
y(E) = a+bE^c
\end{equation}
with $a$, $b$, and $c$ variable parameters. Figure \ref{DSBfit} shows $y(E)$ for single strand breaks and double strand breaks together with the fitted parameters, which are provided in table \ref{table1}.
\begin{table}[h]
\centering
\begin{tabular}{ c | c | c |}
par&SSB&DSB\\\hline
a&1136.3.801&49.656\\
b&-0.1735&0.0569\\
c&-0.945&-0.907\\
\end{tabular}
\caption{\label{table1} Parameter fit for number of single resp. double strand breaks Gy$^{-1}$
 cell$^{-1}$ using a function of the form: $a+bx^c$.}
\end{table}

\item The total relative yield of damage of type D ($Y_D$) was then given by
\begin{equation}
Y_D~=~\sum_{E=E_{min}}^{E_{max}} y_D(E)F(E)
\end{equation}
With F(E) the normalized histogram of deposited energy as a function of energy of the depositing electrons (in MCNPX tally F6:E divided by the total energy). 
\item This repeated for all damage types and volumes with and without nano--particles.
\item The ratio's of the different yields for these volumes provides the relative yield.
\end{enumerate}
\par For reasons of simplicity we chose to concentrate on the yields of single and double strand breaks. The latter being defined as strand breaks on different DNA--helices not more than 10 base pairs apart\cite{Nikjoo1997}.

\section{Results}
\subsection{Enhancement}
\begin{figure}[h] 
\begin{center}
\subfloat[Skin ratio]{\label{Ea}\includegraphics[width=0.4\columnwidth]{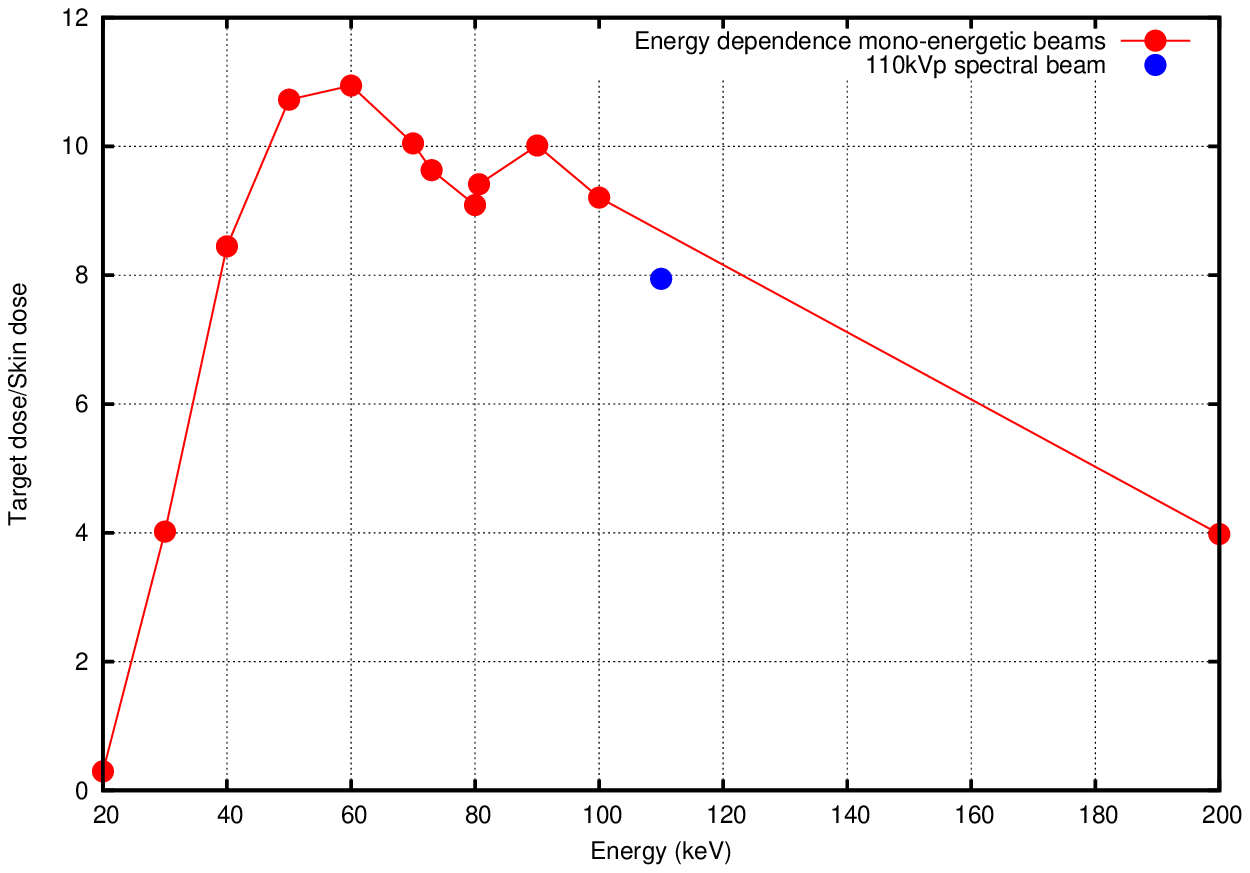}}
\subfloat[OAR Enhancement]{\label{Eoar}\includegraphics[width=0.4\columnwidth]{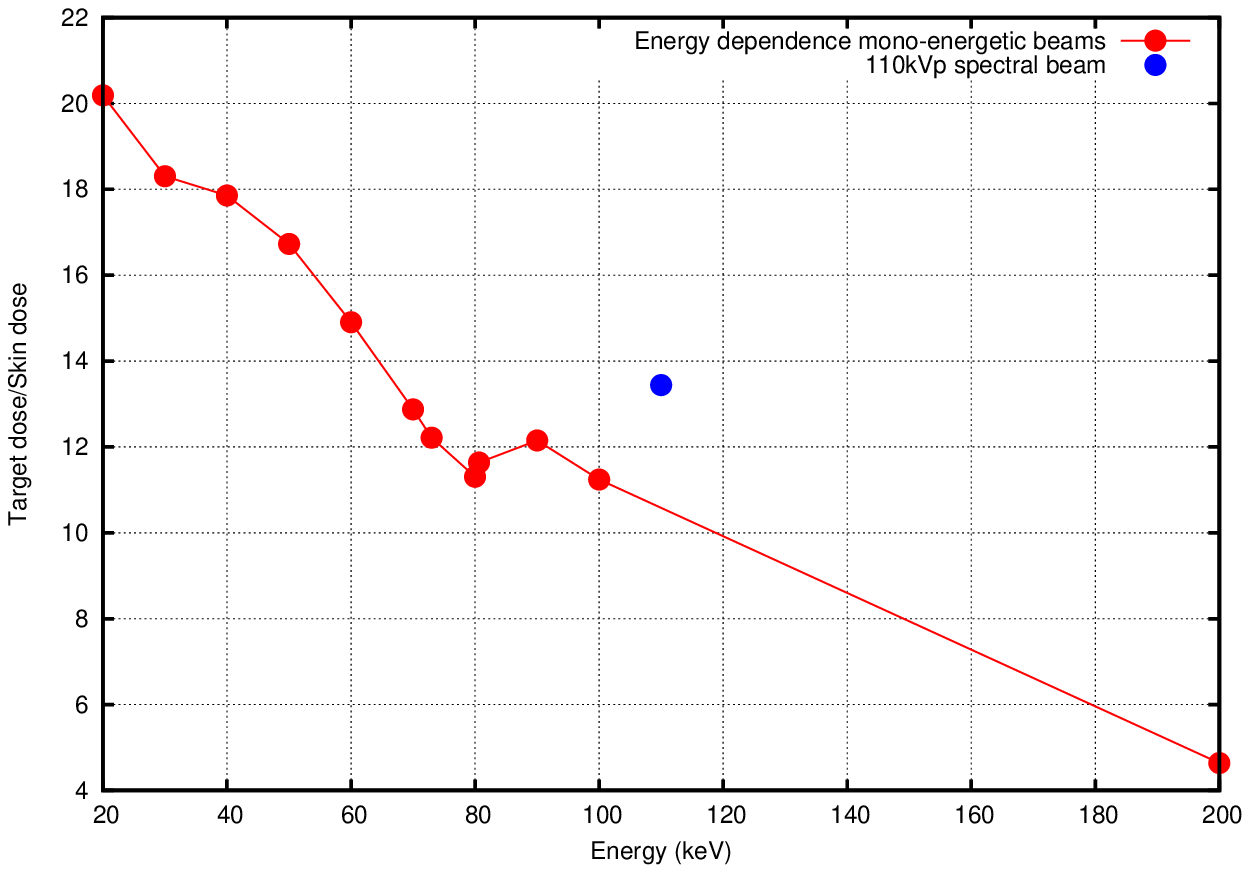}}
\end{center}
\caption{\label{energy1} Enhancement ratio's as function of energy. For
all points error bars are drawn but are too small for visualization,
as all simulations were performed to yield errors smaller than 1\%}
\end{figure}

Figures \ref{Ea} and \ref{Eoar} show the maximal enhancements (i.e. 10\% solution) of the dose in the volume containing GNP (target) compared to the skin and organ at risk structures. The skin ratio shows two local maxima at 60keV and at 90 keV. The OAR, which is positioned 
downstream of the target structure decreases monotonically reflecting a shielding effect due to the increased photon absorption in the 
target. 
\par 
\begin{figure}[h]
\begin{center}
\includegraphics[width=0.7\columnwidth]{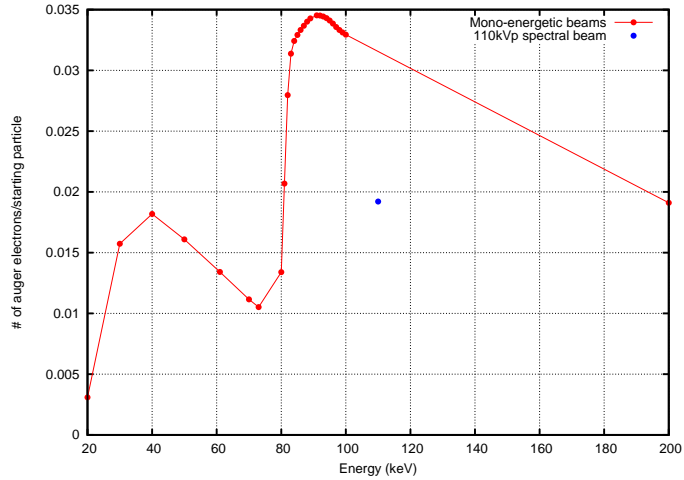}
\end{center}
\caption{\label{auger_numbers} The number of electrons generated from
the different possible channels in the medium containing the gold
concentration as a function of energy for quasi mono--energetic beams. The
point at 110keV represents the result from a bremsstrahlung--spectrum. It
is clear that most of the electrons come from knock--on events, while the
photo--electric electrons are a distant second. Both the PE and Auger
electrons show a maximum around 40keV while an sudden increase is also
noted at the K$_\alpha$--edge. Note that there still is a substantial
contribution of Auger electrons in the bremsstrahlung beam. }
\end{figure}

\begin{figure}[h]
\centering
\includegraphics[width=0.45\columnwidth]{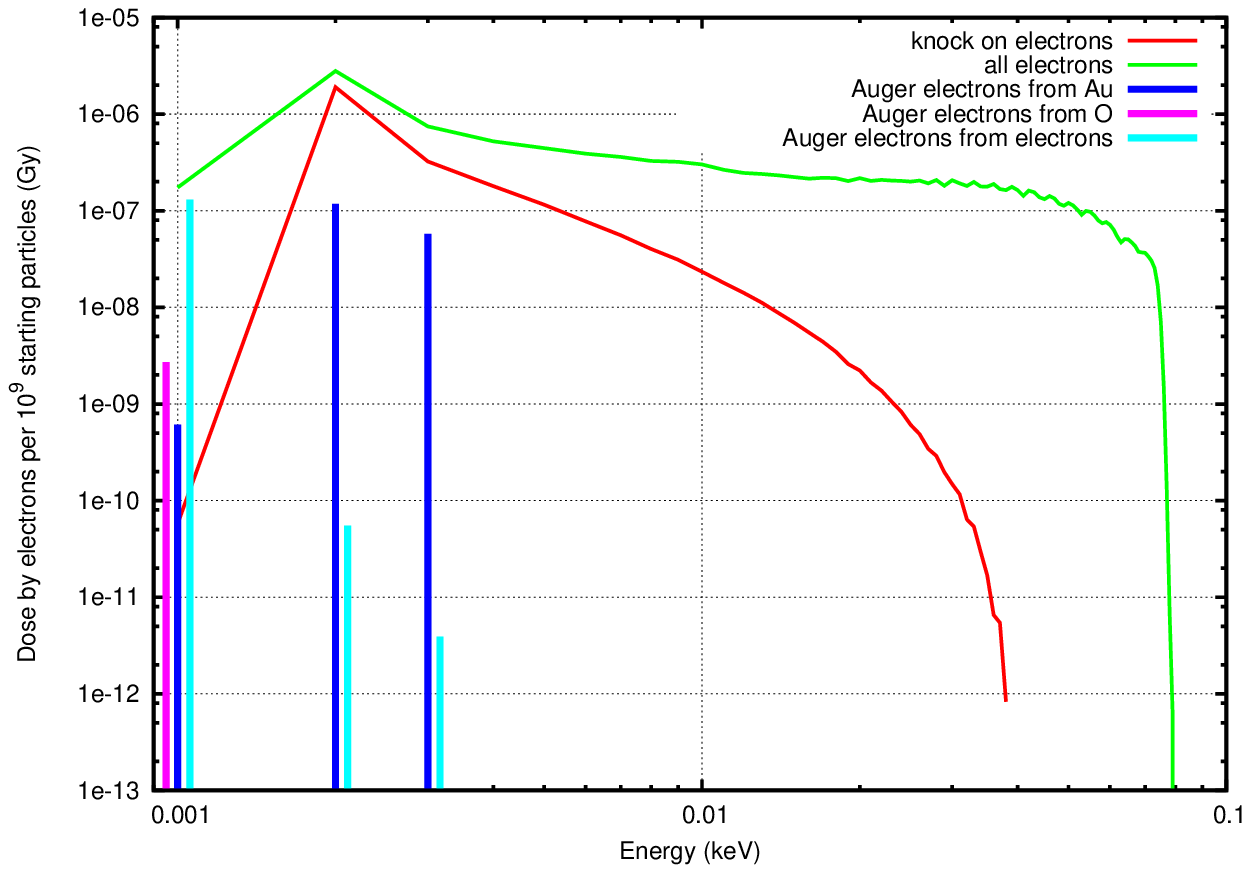}
\includegraphics[width=0.45\columnwidth]{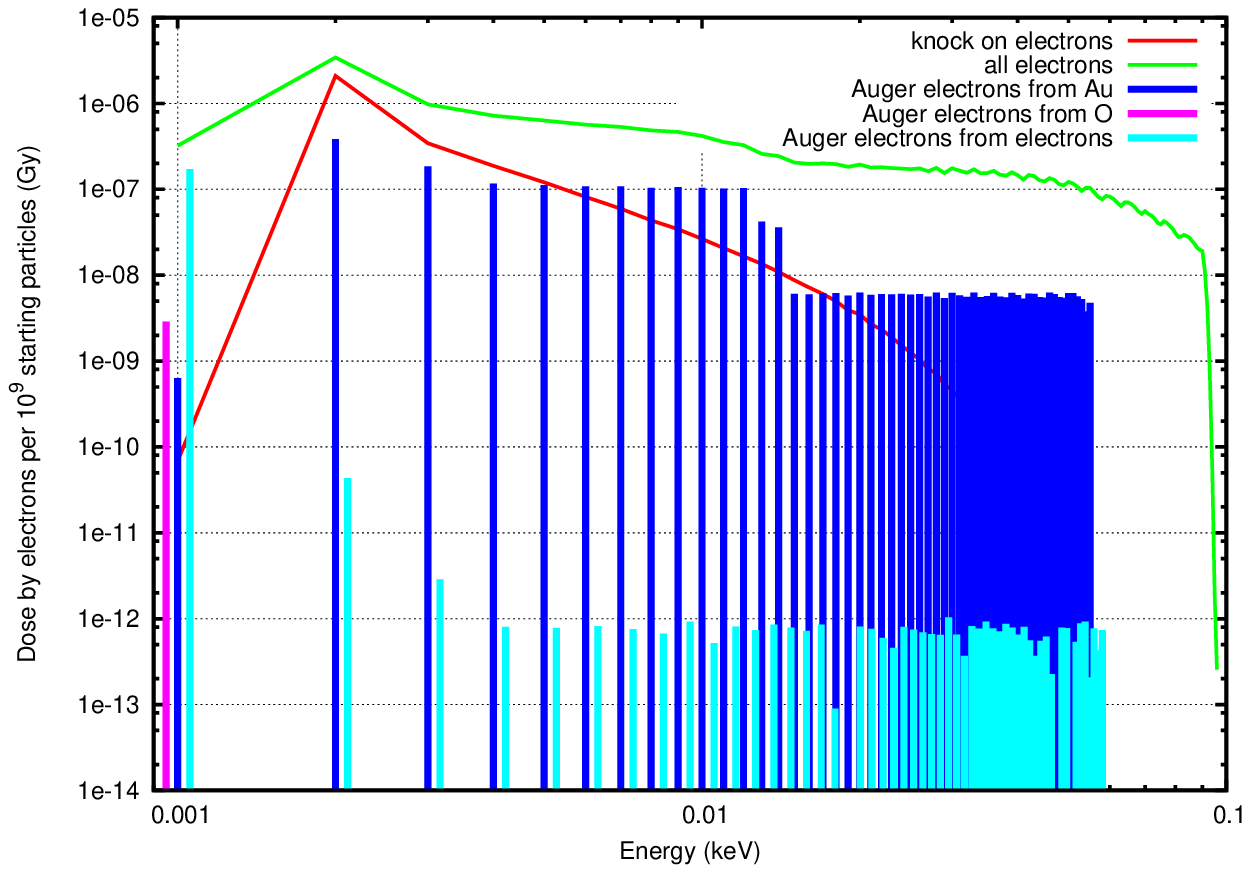}
\caption{\label{auger_energy} Energy spectra of depositing electrons in the activated volume, when using and energy below (left)  the K$_\alpha$--edge of golds,
contrasted with the spectrum resulting from irradiation with an photon beam of 90keV which is above this edge (right).}
\end{figure}

Figure \ref{auger_numbers} counts the number of Auger or Coster--Kr\"onig electrons (AE) generated in the target. The graph reflects the atomic structure in of gold as both L-- and K--edges show their influence leading to local maxima of generated Auger electrons. The highest number generated occurs at 90 keV. However, the energy of the AE in the latter case is much higher than those generated with lower energy. This can be seen in Figure \ref{auger_energy} where the energy deposition of the various electrons is presented. Also note that in both cases the contribution of AE to the energy deposition process remains an order of magnitude lower. The largest part of the dose is deposited by photo--electric (PE) and knock--on electrons. 

\subsection{Concentration}
\begin{figure}[h]
\begin{center}
\includegraphics[width=0.7\textwidth]{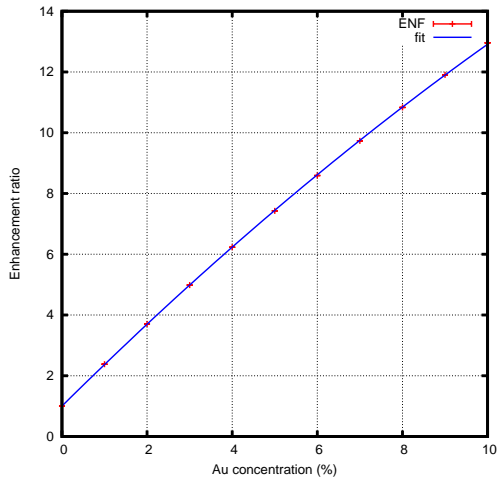}
\end{center}
\caption{\label{conc} Graph depicting the variation  of the
absolute enhancement factor ($E_a$) as a function of the gold
concentration. A second order polynomial is fit to the values using
a $\chi^2$--minimization.}
\end{figure}

Figure \ref{conc} shows the Absolute enhancement ratio as a function
of the concentration of gold in the cell. $E_a$ was fit with a second
order polynomial, where the quadratic factor is small with respect to
the other factors. Showing that the dose enhancement is quasi--linear in the sense that the quadratic term in the
polynomial has a small contribution. 
When the second order polynomial is
denoted as $f(x)~=ax^2 + bx +c$. Then the coefficients are: a= ($-0.0195
\pm 0.0005$), b = ($1.386 \pm 0.004$), and c= ($1.003 \pm 0.003$). Fixing
the $c$--coefficient to 1 gives comparable results.
If a linear fit is performed the slope is given by ($1.26\pm0.02$), a
number which can be used to easily predict the impact of concentration
changes.
\subsection{Biological effect}
\begin{figure}[h]
\begin{center}
\includegraphics[width=0.45\textwidth]{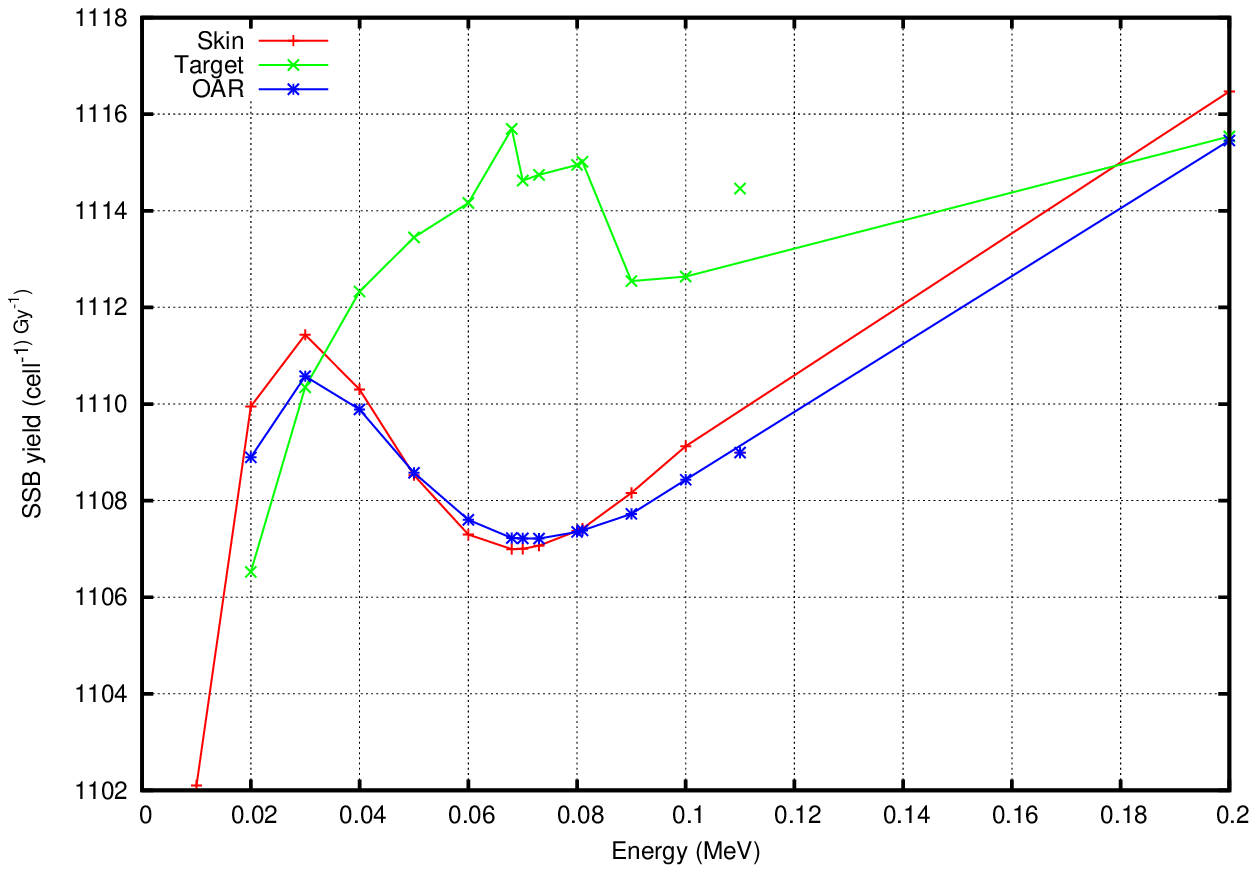}
\includegraphics[width=0.45\textwidth]{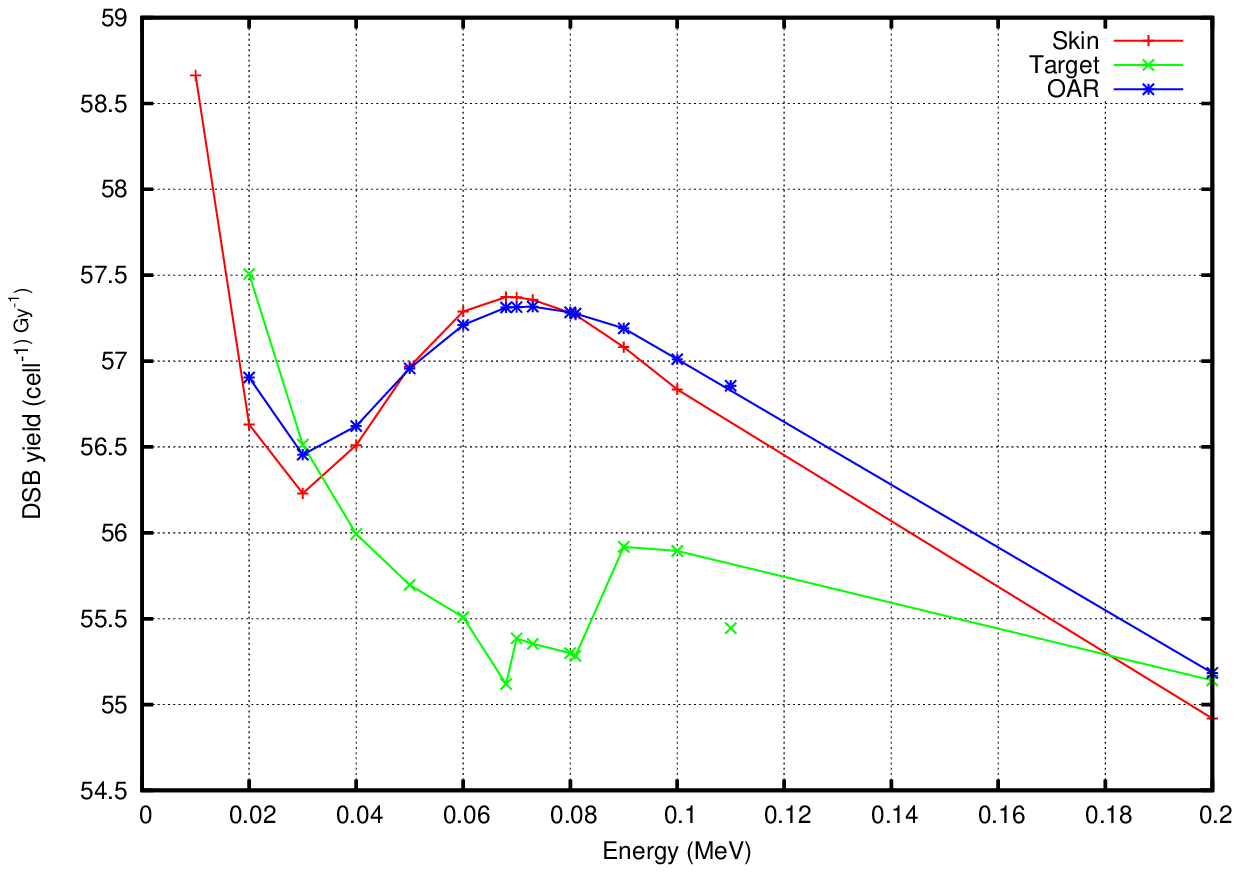}
\end{center}
\caption{\label{RBE} 
Plots for generation of single strand breaks (left) and double strand breaks (right) as a function of energy for a given dose. 
The curves are complementary as the number of SSB's goes down the DSB's go up. In the target volume the atomic structure and
resonant levels generating auger electrons are visible but have a limited impact on the overall result with respect to the 
deposition being of a high or low LET nature. 
}
\end{figure}

Figure \ref{RBE} shows the yield per cell and per Gy of single strand breaks (SSB) and
double strand breaks (DSB) as a function of energy. This for the three volumes under consideration. 
The target volume containing GNP reflects the atomic structure in the calculated yields. One can easily 
identify the when auger electrons are generated as a function of beam energy. The addition generates increases 
in DSB and lowers the contributions of SSB. However, the added tail of high energy electrons 
which mainly affect the DNA through SSB's. It is only for very low energies (10 -- 20keV) that the addition of GNP 
changes the ratio of SSB to DSB to give the dose deposition a more high LET character. For deeper lying structures there 
is no data available for 10 keV as the photons do not penetrate deep enough to generate a meaningful contribution. 
\par
\begin{figure}[h]
\begin{center}
\includegraphics[width=0.7\textwidth]{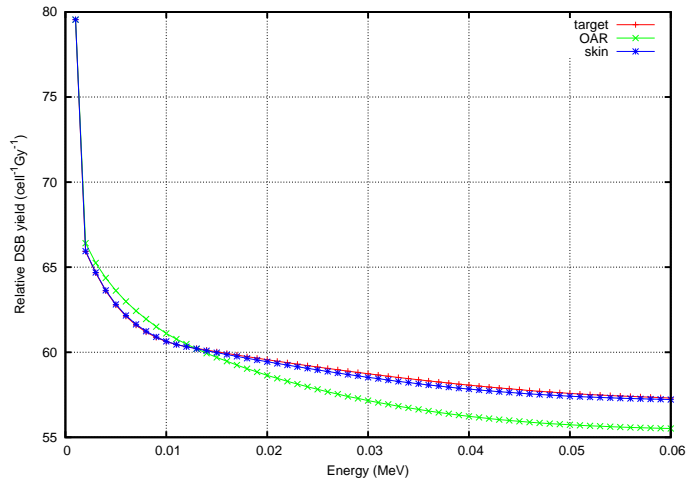}
\end{center}
\caption{\label{Cumul}
Cumulative graph of the contribution to dose through double strand breaks. For each energy bin the contribution is added to the 
higher energies. For the GNP volume the contribution of low energy auger electrons has a relatively high effect. As the higher energy electrons are added the effect diminishes.}
\end{figure}

The spectrum in the target does contain more low energy electrons compared to the OAR's. This advantage is negated by the
additional tail of high energy electrons as illustrated in Fig. \ref{Cumul} where a cumulative representation of the 
dose deposition in double strand breaks is shown for a beam energy of 60kev.

\section{Discussion}
From the data available here, one can conclude that there
exists a necessity to perform full monte carlo simulation based for these
type of treatments. Not only does the enhancement change with energy it is
also imperative to monitor the concentration of the nano--particles during
treatment. Before treatments can be started methodologies need to be
developed to monitor concentrations adequately. The monitoring of the concentration and the 
error of this specific measurement has a direct impact on the dosimetry
as we find a relationship of concentration to dose which is close to
unity. This implies that the uncertainty of the dosimetric planning is
at least equal to the uncertainty of this measurement.
\par
The choice of energy
of the radiation beam for treatment does not seem to be a critical issue, as long as
radiation reaches the intended volume. So, it  might not be necessary
to build expensive high output mono--energetic photon sources.
However, to determine the concentration it might be necessary to do so
albeit with a lower photon flux \cite{ISI:000168216600002, ISI:000229866100003}. A possible alternative for measurement is
the use of the same markers with nano--particles attached, but where the
gold is replaced by a radioactive isotope\cite{Roy2006}. Using methodology developed in
nuclear medicine imaging techniques the concentration and its variation
can then be monitored over time. This approach gives up some of the
advantages as the radioactive particles will deliver a dose in the
manner we are trying to avoid. The dose, however, is only for imaging
purposes and is more limited than when a therapeutic procedure is
attempted. A further approach could be to measure concentration changes
using other nano--particles attached to same targeting agent and use
non--ionizing techniques for visualization\cite{Sun2008}. A drawback
of this technique is that due to the use of different nano--particles,
the uptake and concentration dynamics could be slightly different from
the therapeutically enhanced targeting molecules.
\par
The impact of Auger electrons seems to be limited as they only consist of a small fraction
of the dose depositing electrons. Only at very low energies they seem to have an effect in increasing the 
efficiency of the dose deposition. Using such low energies limits the use to 
superficial tumors or warrants the use of intra--operative techniques or brachytherapy techniques using low energy sources like 
$\mathrm{^{125}I}$ or electronic brachytherapy sources\cite{rivard:4020}.
\par From the data available here it seems that a classical bremsstrahlung source is suited as an investigative tool to assess the 
effectiveness of this approach for deep seated tumors. 
\bibliography{/home/fvdheuve/elsart/generic4}
\bibliographystyle{elsart-num}
\end{document}